\documentclass[pre,twocolumn,showpcs,amsmath,amssymb]{revtex4}

\usepackage{graphicx}
\usepackage{dcolumn}
\usepackage{bm}

\begin{document}

\title{Asymptotic properties of a bold random walk}

\author{Maurizio Serva}

\affiliation{Dipartimento di Ingegneria e Scienze dell'Informazione e Matematica,
Universit\`a dell'Aquila, 67010 L'Aquila, Italy}
\affiliation{Departamento de Biof\'isica e Farmacologia, Universidade 
Federal do Rio Grande do Norte, 59072-970 Natal-RN, Brazil}

\date{\today}

\begin{abstract}
In a recent paper \cite{S} we proposed a non-Markovian random walk model 
with memory of the maximum distance ever reached from the starting point (home).
The behavior of the walker is at variance with respect to the simple symmetric 
random walk (SSRW) only when she is at this maximum distance, where, 
having the choice 
to move either farther or closer, she decides with different probabilities.
If the probability of a forward step is higher then the probability of
a backward step, the walker is bold and her behavior turns out to
be super-diffusive, 
otherwise she is timorous and her behavior turns out to be sub-diffusive.
The scaling behavior vary continuously from sub-diffusive (timorous) 
to super-diffusive (bold) according to a single parameter $\gamma \in R$.
We investigate here the asymptotic properties of the bold case
in the non ballistic region $\gamma \in [0,1/2]$, a problem which was 
left partially unsolved in \cite{S}. 
The exact results proved in this paper require new probabilistic tools 
which rely on the construction of appropriate martingales of the
random walk and its hitting times.
\end{abstract}

\maketitle

 
The appellative anomalous diffusion is associated to a scaling relation
${\it E}[ x^2(t)] \sim t^{2 \nu}$ with $\nu \neq 1/2$.
It may arise in random walks via diverging steps length, 
as in L\'evy flights \cite{L} or via long-range memory as in
self avoiding random walks \cite{APP,TW}. 
Diverging steps length and long-range memory 
are two different ways of violating the necessary conditions for the 
central limit theorem when applied to random walks.

In some cases, the mechanism which gives origin to anomalous scaling can 
be different, special deterministic or random
environments \cite{VS,CS} or multi-particle 
interactions \cite{LB}.
Moreover, diffusion can be strongly anomalous 
(${\it E}[ |x(t)|^q] \sim t^{q \nu}$ with $\nu$ depending on $q$) 
in complex systems \cite{PV,CM,AC}.

There is a very large number of phenomena which exhibit anomalous diffusion
as well a variety of models which have been used to describe them,
(for a review of both see \cite{BG,BH,MK1,MK2,RKS}). 
Nevertheless,
exact solutions of non-trivial models with memory are quite scarce
\cite{ST,BF,SK,BSS,BR,DAB,BP}.
Motivated by this lack of exact solutions,
we presented in \cite{S} a model which is exactly treatable although
genuinely non-Markovian.
The model shows anomalous scaling which can be sub-diffusive,  
super-diffusive and also ballistic 
according to a single parameter $\gamma \in R$.

It is the aim of this work to investigate here the asymptotic properties 
in the range $\gamma \in [0,1/2]$, a problem which was left partially 
unsolved in \cite{S}. This range corresponds to
the a non ballistic super-diffusive behavior except at the two extreme
were it is ordinary SSRW ($\gamma=0$) and ballistic ($\gamma=1/2$).

The model, as defined in \cite{S}, is one-dimensional, steps all have the 
same unitary length, time is discrete and the walker can only move left
or right at any time step.
The behavior of the random walker is modified 
with respect to SSRW only when she is
at the maximum distance ever reached from her starting point (home).
In this case, she decides with different probabilities
to make a step forward (going farther from home) or a step backward
(going closer to home). 
 
More precisely, the model is the following: 
the walker starts from home ($x(0)=0$), then, at any time she can make a 
(unitary length) step to the right or to the left, i.e.
\begin{equation}
x(t+1)=x(t)+\sigma(t)
\label{x}
\end{equation}
with $\sigma(t)= \pm 1$. Then, let us define 
\begin{equation}
z(t)= \max_{0 \le s \le t} |x(s)|,
\label{y}
\end{equation}
which is the maximum distance from home she ever attained.

We assume that the walker has no preference for the direction 
of the first step ($\sigma(0)= \pm 1$ with equal probability),
as well she has no preference when she is not at the maximum distance
($\sigma(t)= \pm 1$ with equal probability if $|x(t)| <z(t) \,$).
On the contrary, when she is at the maximum distance ($|x(t)|=z(t)$), 
she can chose to step away from the origin with probability
$p(z(t))$ or toward the origin with probability $1-p(z(t))$,  i.e.
$\sigma(t)= {\rm sign}(x(t))$ with probability $p(z(t))$ and 
$\sigma(t)= -{\rm sign}(x(t))$ with probability $1-p(z(t))$.
It  is assumed that the probability $p(z)$ depends on $z$ according to 
\begin{equation}
p(z)= \frac{z^\gamma}{1+z^\gamma},
\label{p}
\end{equation}
where $\gamma \in R$.

Therefore, the walker performs a simple symmetric random walk (SSRW) when 
$|x(t)|<z(t)$, however, when she is at the maximum distance from home 
($|x(t)|=z(t)$),
she boldly prefers to move farther if $\gamma > 0$ or timorously prefers 
to move closer if $\gamma < 0$.
Moreover, if her attitude is neutral ($\gamma=0$),
the model is globally SSRW since 
$p(z)=1/2$ and all steps are all and always equally probable. 

The main results of both present paper and paper \cite{S}, 
concerning the asymptotic behavior of $z(t)$, 
are summarized by the following limits which hold
in probability for large times ($t \to \infty$):
\begin{itemize}
\item
$ \gamma \in (-\infty,0)$: $\; z(t)/t^\nu \to (1/2\nu)^{\nu} $ where
the scaling exponent is $\nu=1/(2-\gamma)$,
\item
$ \gamma = 0 $: $\; z(t)/t^{1/2} \to 1/T^{1/2} $ where $T$ is 
a random variable described below,
\item
$ \gamma \in (0,1/2) $: $\; z(t)/t^\nu \to (1/2\nu)^{2\nu}/L^{\nu} \,$ 
where $\nu=1/(2-2\gamma)$ and $L$ is a L\'evy variable,
\item
$ \gamma = 1/2 $: $\; z(t)/t \to 1/(4L+1)$ where $L$ is 
the same L\'evy variable, 
\item
$ \gamma \in (1/2,\infty) $: $\; z(t)/t \to 1$.
\end{itemize}

We will show in the next that the
Laplace transforms of probability densities of $L$ and $T$ are:

\begin{equation}
{\it E}[ e^{-\lambda L}]= 
e^{-\sqrt{2\lambda}}, \quad \quad
{\it E}[ e^{-\lambda T}]= 
\frac{1}{\cosh(\sqrt{2\lambda})}.
\label{LT}
\end{equation}
Accordingly, $L$ is a L\'evy variable with parameter 1, and it corresponds 
to the first hitting time of a barrier in 1 by a continuous Brownian motion 
with unitary diffusion constant, starting in 0.
The  variable $T$ is the first exit time from the interval $[-1,1]$ 
by the same Brownian motion. Explicitly, the probability density of $L$ is 
\begin{equation}
\rho(L)= \sqrt{\frac{1}{2\pi L^3}} \,
\exp{\!\left(\!-\frac{1}{2L}\right)}.
\label{rL}
\end{equation}
Thus, in the range $\gamma \in (0,1/2) $, one has
\begin{equation}
\lim_{t \to \infty }{\it E}
\left[ \left(\frac{z(t)}{t^\nu}\right)^q\right]= 
\left(\frac{1}{2\nu^2}\right)^{q\nu} \, 
\frac{ \Gamma(q\nu+1/2)}{\Gamma(1/2)},
\label{Lq}
\end{equation}
where $\Gamma$ is the gamma function, $q$ is a real
positive constant and $\nu=1/(2-2\gamma)$.

The sub-diffusive range $ \gamma \in (-\infty,0)$ and 
the ballistic range $ \gamma \in (1/2,\infty)$,
where already fully solved in \cite{S} where both $\nu$ and 
the exact (constant) values of the limits were determined.

In this paper we focus on the remaining range $ \gamma \in [0,1/2]$
which in \cite{S} was solved only for what concerns the scaling exponent $\nu$.

For $ \gamma = 0$ the model reduces to SSRW, all of which is already known.
Nevertheless, for sake of comparison, we re-obtain here already known results 
by our method.
The range $ \gamma \in (0,1/2)$ corresponds to a non-ballistic
super-diffusive behavior ($1/2 < \nu < 1 $). We prove here that
the ratio $z(t)/t^\nu$ is distributed as $(1/2\nu)^{2\nu}/L^{\nu}$
for large times.
For $ \gamma = 1/2$ one has a ballistic behavior, but the large time limit of
$z(t)/t$ is not 1 as in the range $ \gamma \in (1/2,\infty)$.
We show here that it is distributed as $1/(4L+1)$,
meaning that the walker spends only a finite but random fraction of her
time moving linearly away from home.

%
Let us now outline our mathematical approach.
Trajectories are decomposed in active journeys and lazy journeys. 

The lazy journey starts at time $t$ when a walker, which is on a maximum 
$x(t)=z(t)$ or $x(t)=- z(t)$, leaves it (first step)
and continues for $m$ time steps until she reaches again one of the two maxima. 
The total number of time steps of this journey is $1+m$ where $m$ is the random 
time necessary to hit the frontier of the interval 
$[-z(t),z(t)]$ starting from one of the two positions $z(t)-1$ or $-z(t)+1$.
During all the lazy journey the maximum remains the same ($z(t+m+1)=z(t)$)
and, very importantly, the $m$ steps of the walk necessary for
hitting the frontier are those of a SSRW. Notice, in fact, that
all the $m$ steps are made choosing the direction with equal
probability.

The active journey starts at the time $t+m+1$ when the walker 
arrives on a maximum and it has a duration of $n$ time steps
which she makes remaining on a maximal position. 
The $n$ steps are all made in the opposite direction with respect 
to the origin.
In numbers:  $|x(t+m+1+s)|=z(t+m+1+s)= z(t)+s$ for $0 \le s \le n$ 
while  $|x(t+m+n+2)|<z(t+m+1+n)$, this last being the first step 
of a new lazy journey. 
The active journey has a minimum duration of zero time steps 
($n=0$ when the walker immediately leaves the maximum after being arrived).
During the active journey the maximum increases of $n$.

A cycle journey, starting in a position $|x(t)|=z(t)$, is composed by a 
lazy journey followed by an active journey,
its duration is $1+m+n$ and the maximum increases of $n$. 

Notice that $m=m(z)$ is a random variable whose 
distribution only depends on $z$. In fact,
$m(z)$ is the SSRW first hitting time of one of the barriers
$z$ or $-z$ starting from position $x=z-1$ or $x=-z+1$.
On the contrary, the distribution of $n=n(z)$ depends
both on $z$ and $\gamma$ through $p(z)$.

Let us indicate by $k$ (not to be not confused time $t$)
the progressive integer number identifying cycle journeys, 
each composed by a lazy journey followed by an active journey.
Also, let us indicate by $z(k)$ the value of the maximum when the cycle 
journey number $k+1$ starts.

Then, the time $t$ is linked to the progressive number $k$ by the 
stochastic relation
\begin{equation}
t(k+1)= t(k)+ 1+ m(z(k))+n(z(k)),
\label{tk}
\end{equation}
while the value of the maximum is given by
\begin{equation}
z(k+1)= z(k)+n(z(k)),
\label{yk}
\end{equation}
where $m(z(k))$ and $n(z(k))$ are all independent random variables.

For the sake of completeness let us also write down the initial
condition. At the start ($x(0)=z(0)=0$) the walker moves left or right
so that $x(0)=\pm 1$ and $z(1)=1$. Then, starting from the maximum $z(1)=1$,
she begins an active journey (which can also
be of $n(1)=0$ steps if she immediately steps back to the origin)
so that:
\begin{equation}
t(1)=z(1)= 1+n(1).
\label{ty1}
\end{equation}

In principle one should simply solve the two equations
(\ref{tk},\ref{yk}) with initial condition (\ref{ty1})
in order to obtain the scaling behavior of $z(t)$. 
Obviously, this asks for some work
since we need to characterize probabilistically $m(z)$ and $n(z)$.


Let us start with $m(z)$ which by definition is the SSRW exit time from 
the interval $[-z,z]$ starting in $z-1$ or $-z+1$. 
By translational invariance, $m(z)$ can be also considered as the SSRW 
exit time from the interval $[-2z+1,1]$ starting in 0.

The third Wald identity, when applied to SSRW trajectories
$w(s)$ starting in $w(0)=0$, states that for any stopping time $\tau$

\begin{equation}
{\it E}\left[ \frac{e^{\theta w(\tau) } }
{(\cosh(\theta))^\tau} \right]=1.
\label{b1}
\end{equation}
Considered that 
$e^{\theta w(s)}/(\cosh(\theta))^s$ is a martingale, 
this equality is a simple consequence of the strong Markov property.
The above equation (\ref{b1}) also holds if $\theta$ is replaced by $-\theta$
so that
\begin{equation}
{\it E}\left[ \frac{A \, e^{\theta w(\tau) } + (1-A) \, e^{-\theta w(\tau)} }
{(\cosh(\theta))^\tau} \right]=1
\label{b2}
\end{equation}
for any real $A$. 
 
Suppose $a < 0 <b$ and assume that that $\tau=\tau(a,b)$ is the first exit time 
of $w(s)$ from the interval $[a, b]$ so that $w(\tau)=a$ or $w(\tau)=b$. 
One can chose the real constant $A$ in oder that the numerator in (\ref{b2})
has the same value in $w(\tau)=a$ and $w(\tau)=b$ obtaining:

\begin{equation}
{\it E}[ (\cosh(\theta))^{-\tau}]= 
\frac{\cosh(\theta c)}{\cosh(\theta  d)},
\label{b3}
\end{equation}
where $c=(a+b)/2$ and $d=(b-a)/2$. 

Then, having defined $\lambda= \ln (\cosh(\theta))$,
one can rewrite the above equality as a Laplace transform of
the distribution of the stopping time $\tau=\tau(a,b)$

\begin{equation}
{\it E}[ e^{-\lambda \tau}]= 
\frac{\cosh(\theta(\lambda) \, c)}{\cosh(\theta(\lambda) \,  d)},
\label{b4}
\end{equation}
where $\theta(\lambda)= \ln(e^{\lambda} + \sqrt{e^{2\lambda}-1})$.

We  simply use $a=-2z+1$ and $b=1$ so that $c=1-z$ and $d=z$. 
Thus, the Laplace transform
of the probability density of the exit time $m(z)$ is

\begin{equation}
{\it E}[ e^{-\lambda m(z)}]= 
\frac{\cosh[\theta(\lambda) \, (z-1)]}{\cosh[\theta(\lambda) \, z]}.
\label{b5}
\end{equation}

From (\ref{b5}) one can derive the
expected values of all powers of $m(z)$. 
For large values of $z$ one finds
${\it E}[ m(z)] \simeq 2z$ and ${\it E}[ m^2(z)] \simeq (8/3)z^3$,
which implies that the standard deviation is
$\sigma_{m(z)} \simeq (8/3)^{1/2} z^{3/2}$. 
All these quantities diverge for large values of $z$.

In the limit of large $z$, nevertheless, the Laplace transform remains
finite and well defined; one has, in fact,
${\it E}[ e^{-\lambda m(z)}] \to e^{-\theta(\lambda)}$.
Moreover, for small values of $\lambda$ one as that 
$\theta(\lambda) \simeq \sqrt{2 \lambda}$, meaning that
the probability density of $m(z)$ is substantially
a truncated L\'evy density.

Let us now evaluate the probability $ \pi_\gamma\,(n|z)$ 
that the walker makes at least $n(z)=n$ steps during the active journey, 
i.e. $\pi_\gamma\,(n|z)={\rm prob}\,(n(z) \ge n)$.
Straightforwardly:
\begin{equation}
\pi_\gamma\,(n|z)= \prod_{s=0}^{n-1} p(z+s), 
\label{a1}
\end{equation}
where $ p(z+s)= (z+s)^\gamma/(1+(z+s)^\gamma)$.

At variance with $m(z)$, the variable $n(z)$ depends on $\gamma$.
In this paper we focus on the range $\gamma \in [0,1/2]$
and, in order to describe the probabilistic behavior of $n(z)$,
we have to distinguish two different sub-ranges. 

The first is $\gamma=0$, for this value (ordinary 
SSRW) one has $p(z)= 1/2$ and $\pi_\gamma\,(n|z)= (1/2)^n $. 
Accordingly, ${\it E}[ n(z) ] =1$ and all averages ${\it E}[ n(z)^\delta]$ 
are finite and they are independent from $z$ for any positive $\delta$.

The second case corresponds to the range $\gamma \in (0,1/2]$,
included in $(0,1)$, which, in turn, can be treated at once.
We directly obtain from (\ref{a1}), 
\begin{equation}
[p(z)]^{n} \le \pi_\gamma (n|z) \le  [p(z+n-1)]^{n}.
\label{a2}
\end{equation}
In fact, $\nu$ being positive,
$p(z)$ is the smallest among the elements of the product
and $p(z+n-1)$ the largest.

Then assume $n=I[\beta z^\gamma]$ (the integer part)
where $\beta$ is real and strictly positive.
The inequality (\ref{a2}) rewrites 
\begin{equation}
[p(z)]^{I[\beta z^\gamma]} \le \pi_\gamma (n|z)
\le  [p(z+I[\beta z^\gamma])]^{I[\beta z^\gamma]}.
\label{a3}
\end{equation}

Then, since $\gamma \in (0,1)$, one gets that the limit for 
$z \to \infty$ of both bounds is $e^{-\beta}$.
Given that $ \pi_\gamma (n|z) = {\rm prob}\,(n(z) \ge I[\beta z^\gamma])$,
one finally has
\begin{equation}
{\rm prob}\,(n(z) \ge \beta z^\gamma) \simeq e^{-\beta}.
\label{a4}
\end{equation}

The approximated equality (\ref{a4}) means that for large values of $z$
the limit $n(z) /z^\gamma \to \xi$ holds
where $\xi$ is a random variable distributed according
to an unitary exponential probability density.
Since $n(z) \simeq \xi z^\gamma$, one can easily compute
${\it E}[ n(z) ] \simeq z^\gamma$
and ${\it E}[ n(z)^\delta] \sim  z^{\delta\gamma} $ for any positive $\delta$. 


Summarizing, the relation $ {\it E}[ n(z)^\delta] \sim  z^{\delta\gamma}$ 
holds in all range $\gamma \in [0,1)$ and, thus, in the range $[0,1/2]$.

Let us consider again equation (\ref{yk}), 
one has for any positive $\alpha$
\begin{equation}
{\it E}[z(k+1)^{\alpha}]
\simeq  {\it E}[z(k)^{\alpha}] +\alpha {\it E}[z(k)^{\alpha-1+\gamma}],
\label{y1}
\end{equation}
where the omitted terms are of lower order in $z(k)$ since  
the conditional expectation of $n(z(k))^\delta$ given $z(k)$
satisfies $ {\it E}[ n(z(k))^\delta] \sim  
z(k)^{\delta\gamma} \ll z(k)^{\delta}$. 
Choosing $\alpha=1-\gamma$, we immediately obtain by integration 
${\it E}[ z(k)^{1-\gamma} ] \simeq (1-\gamma) k$. Then, 
choosing  $\alpha=l(1-\gamma)$, we get by iteration
${\it E}[ z(k)^{l(1-\gamma)} ] \simeq (1-\gamma)^l k^l$ 
where $l$ is any positive integer number.

Thus,  ${\it E}[ z(k)^{l(1-\gamma)} ] \simeq {\it E}[ z(k) ]^{l(1-\gamma)} 
\simeq (1-\gamma)^l k^l$ which implies that the relation
\begin{equation}
z(k) \simeq (1-\gamma)^{1/(1-\gamma)} \, k^{1/(1-\gamma)} 
\label{zkf}
\end{equation}
holds deterministically in the range $\gamma \in [0,1)$, i.e. the 
large $k$ limit of the ratio of the two sides of (\ref{zkf}) is one.

On the other hand, from equation (\ref{tk}) we have 
by a direct sum
\begin{equation}
t(k)= z(k)+k^2 L(k) +k-1,
\label{tkf}
\end{equation}
where we have defined
\begin{equation}
L(k)= \frac{1}{k^2} \sum_{i=1}^{k-1} m(z(i)).
\label{c1}
\end{equation}
Then, we can use (\ref{b5}) and straightforwardly obtain
\begin{equation}
{\it E}[ e^{-\lambda L(k)}]= \prod_{i=1}^{k-1}
\frac{\cosh[\theta(\lambda/k^2) \, (z(i)-1)]}
{\cosh[\theta(\lambda/k^2) \, z(i)]},
\label{c2}
\end{equation}
where $z(i)$ is given by (\ref{zkf}).
This expression can be rewritten as 
\begin{equation}
{\it E}[ e^{-\lambda L(k)}]= 
\left[ e^{-\theta(\lambda/k^2) } \right]^{k-1} R(k),
\label{c4}
\end{equation}
where
\begin{equation}
R(k)=
\prod_{i=1}^{k-1}
\frac{1+e^{-2\theta(\lambda/k^2) \, (z(i)-1)}}
{1+e^{-2\theta(\lambda/k^2) \, z(i)}}.
\label{c5}
\end{equation}
It is easy to check that in the limit of large $k$ one has
$ \left[ e^{-\theta(\lambda/k^2) } \right]^{k-1} 
\to e^{-\sqrt{2\lambda}}$
Moreover, some lengthy but straightforward calculations lead to
$R(k) \to 1$ for $\gamma \in (0,1)$, while for $\gamma=0$ they lead to
$R(k) \to 2/(1+e^{-2\sqrt{2\lambda}} ) $.

In conclusion, for $\gamma \in (0,1)$,
\begin{equation}
\lim_{k \to \infty} {\it E}[ e^{-\lambda L(k)}]= e^{-\sqrt{2\lambda}},
\label{c9}
\end{equation}
which implies that $L= \lim_{k \to \infty} L(k)$ is a (parameter=1)
L\'evy variable which has an infinite expectation.
Another way to see this result
is to consider it as a direct consequence of the generalized central
limit for leptokurtic variables \cite{L}.

On the other hand, for $\gamma=0$,
\begin{equation}
\lim_{k \to \infty} {\it E}[ e^{-\lambda L(k)}]= 
\frac{1}{\cosh(\sqrt{2\lambda})},
\label{c10}
\end{equation}
which corresponds to a variable $T= \lim_{k \to \infty} L(k)$
with finite expectation and standard deviation
(1 and $\sqrt{2/3}$ respectively).
This variable is the exit time from the interval $[-1,1]$
of a continuous Brownian motion, with unitary variance, starting in 0
(see, for example, \cite{BS}, page 212).
Since $\lim_{k \to \infty} L(k)$ equals $L$ for $\gamma \in (0,1/2]$
and it equals $T$ for $\gamma =0$, the above two relations
(\ref{c9}) and (\ref{c10}) coincide with (\ref{LT}).

Now, consider equation (\ref{tkf}) for large values of $k$ and 
take into account that $\lim_{k \to \infty} L(k)=T$ for $\gamma=0$
and that $\lim_{k \to \infty} L(k)=L$ for $\gamma \in (0,1/2]$.
Also taking into account (\ref{zkf}), one has the asymptotic
relations: $t(k) \simeq k^2 T $ for $ \gamma = 0 $,
$t(k) \simeq k^2 L $ for $\gamma \in (0,1/2)$
and $t(k) \simeq z(k)+k^2 L= k^2/4+k^2 L $ for $\gamma=1/2$.
These relations are obtained neglecting terms of lower order 
with respect to $k^2$.

Finally, taking again into account (\ref{zkf}), 
one obtains for large times (which imply large $k$): 
$\; z(t)/t^{1/2} \to 1/T^{1/2} $ for $ \gamma = 0 $,
$\; z(t)/t^\nu \to (1/2\nu)^{2\nu}/L^{\nu} \,$ 
where $\nu=1/(2-2\gamma)$ for $ \gamma \in (0,1/2) $ and 
$\; z(t)/t \to 1/(4L+1)$ for  $\gamma = 1/2 $.
These results complete the characterization of
the asymptotic behavior of $z(t)$ initiated in \cite{S}.

\bigskip
The author warmly thanks Michele Pasquini, Eudenilson Lins de Albuquerque,
Angelo Vulpiani and Umberto Laino Fulco which contributed
with many discussion and suggestions.

\end{document}